\documentclass{aa}
\usepackage{graphics}
\begin{document}
\thesaurus{ 
	3(08.05.1; 
        09.04.1; 
	09.08.1; 
	09.09.1 N159-5; 
	09.11.1; 
	11.13.1)  
	} 

\def\frac{$''$\hspace*{-.1cm}}
\def\deg{$^{\circ}$\hspace*{-.1cm}}
\def\min{$'$\hspace*{-.1cm}}
\def\h2{H\,{\sc ii}}
\def\hi{H\,{\sc i}}
\def\hb{H$\beta$}
\def\ha{H$\alpha$}
\def\hd{H$\delta$}
\def\heii{He\,{\sc ii}}
\def\hg{H$\gamma$}
\def\sii{[S\,{\sc ii}]}
\def\siii{[S\,{\sc iii}]}
\def\oiii{[O\,{\sc iii}]}
\def\oii{[O\,{\sc ii}]}
\def\hei{He\,{\sc i}}
\def\sm{$M_{\odot}$}
\def\sl{$L_{\odot}$}
\def\ab{$\sim$}
\def\x{$\times$}
\def\sec{s$^{-1}$}
\def\cm2{cm$^{-2}$}
\def\mcube{$^{-3}$}
\def\lam{$\lambda$}

\title{
The ``Papillon'' nebula: a compact \h2 blob in the LMC resolved 
   by {\it HST}\,\thanks{Based on observations with 
   the NASA/ESA Hubble Space Telescope obtained at the Space Telescope 
   Science Institute, which is operated by the Association of Universities 
   for Research in Astronomy, Inc., under NASA contract 
   NAS\,5-26555.}
}

\offprints{M. Heydari-Malayeri, heydari@obspm.fr}

\date{Received 11 June 1999 / Accepted 12 July 1999}

\titlerunning{LMC N\,159-5}
\authorrunning{Heydari-Malayeri et al.}

\author{M. Heydari-Malayeri\inst{1} \and M.R. Rosa\inst{2,}\,\thanks
    {Affiliated to the Astrophysics Division, Space Science 
    Department of the European Space Agency.} \and V. Charmandaris\inst{1} \and
    L. Deharveng\inst{3} \and H. Zinnecker\inst{4}
 }

\institute{{\sc demirm}, Observatoire de Paris, 61 Avenue de l'Observatoire, 
F-75014 Paris, France \and
Space Telescope European Coordinating Facility, European 
Southern Observatory, Karl-Schwarzschild-Strasse-2, D-85748 Garching bei 
M\"unchen, Germany \and 
Observatoire de Marseille, 2 Place Le Verrier, 
F-13248 Marseille Cedex 4, France \and
Astrophysikalisches Institut Potsdam, An der Sternwarte 16, 
D-14482 Potsdam, Germany 
} 

\maketitle

\begin{abstract}
We present high spatial resolution {\it HST} imaging of the LMC
compact \h2\, region N159-5.  This high excitation blob is revealed to
be a {\it ``papillon''} or butterfly-shaped ionized nebula with the
``wings'' separated by \ab\,2\frac.3 (0.6 pc).  Two subarcsecond
features resembling a ``smoke ring'' and a ``globule'' are detected in
the wings, the origin of which is briefly discussed.  N159-5 may
represent a new type of \h2\, region in the Magellanic Clouds
overlooked so far because of insufficient spatial resolution.  Our
images also show a strikingly turbulent medium around the {\it
Papillon} in the giant \h2\, region N\,159, which manifests itself by
a large number of subarcsecond filaments, arcs, ridges, and fronts
carved in the ionized gas by the stellar winds from massive stars in
the N\,159 complex.
 

\end{abstract}

\section{Introduction}

The formation process of massive stars is still a largely unsolved
problem.  Although it is believed that stars generally originate from
the collapse and subsequent accretion of clumps within molecular
clouds (Palla \& Stahler \cite{palla}), this model cannot explain the
formation of stars beyond \ab\,10 \sm\, (Bonnell et
al. \cite{bon}). The strong radiation pressure of massive stars can
halt the infall of matter limiting the mass of the star (Yorke \&
Kr\"ugel \cite{yor}, Wolfire \& Cassinelli \cite{wol}, Beech \&
Mitalas \cite{bee}), while a large fraction of the infalling material
may as well be deflected into bipolar outflows by processes which we
do not yet know in detail (Churchwell \cite{church}).

Moreover, since the evolutionary time scales of massive stars are
comparatively short, these stars are believed to enter the main
sequence while still embedded in their parent molecular clouds (Yorke
\& Kr\"ugel \cite{yor}, Shu et al. \cite{shu}, Palla \& Stahler
\cite{pal}, Beech \& Mitalas \cite{bee}, Bernasconi \& Maeder
\cite{bern}). This means  that massive stars may already experience
significant mass loss and subsequent evolution while still accreting
mass from the parental cloud. 

In order to understand the formation of massive stars it is therefore
necessary to study them at the earliest phases where they can be reached 
through the enshrouding material at different wavelengths. While
high-resolution radio continuum observations allow the investigation of
ultracompact \h2 regions (Churchwell \cite{chur}) formed around
newborn massive stars, high angular resolution observations in
ultraviolet, visible, and infrared wavelengths (Walborn \& Fitzpatrick
\cite{wal}, Walborn et al. \cite{walb}, Schaerer \& de Koter
\cite{sch}, Hanson et al. \cite{hanson}) are necessary to access 
accurate physical parameters of these stars and then evaluate their
states of evolution.  

Our search for the youngest massive stars in the Magellanic Clouds
started almost two decades ago on the basis of ground-based
observations. This led to the discovery of a distinct and very rare
class of \h2 regions, that we called high-excitation compact
\h2 ``blobs'' (HEBs). The blob in N\,159, which is the subject of 
this paper, was the prototype of this category of nebulae 
(Heydari-Malayeri \& Testor 1982). So far only four other  
HEBs have been found in the LMC: N\,160A1, N\,160A2,
N\,83B-1, and N\,11A (Heydari-Malayeri \& Testor 1983, 1985, 1986,
Heydari-Malayeri et al.\,1990) and two more in the SMC: N\,88A and N\,81
(Testor \& Pakull 1985, Heydari-Malayeri et al. 1988). 
To further improve our understanding of those compact \h2\, 
regions and overcome the difficulties related to their small size, we
used the superior resolving power of {\it HST} to image N\,81 and N\,88A,
in the SMC, as well as N\,159-5 in the LMC. The analysis and
discussion for the first two objects was presented by Heydari-Malayeri
et al. (\cite{hey99a},\cite{hey99b}, hereafter Papers I and II
respectively).

In the present paper we study our third {\it HST} target, the LMC blob
N\,159-5. This object lies in the \h2\, complex N\,159 (Henize
\cite{hen}), situated some 30\min\, (\ab\,500 pc) south of 30 Dor.
N\,159 is associated with one of the most important concentrations of
molecular gas in the LMC (Johansson et al. \cite{joh} and references
therein) and contains several signposts of ongoing star formation
(cocoon stars, IR sources, masers).  N\,159-5 is the name given by
Heydari-Malayeri \& Testor (\cite{hey82}) to a compact \h2\, region of
size \ab\,6\frac\, (1.5 pc) with high excitation (\oiii/\hb\,=\,8) and
suffering a considerable extinction of $A_{V}$\,=\,5 mag as derived
from \hb\, and radio continuum (Heydari-Malayeri \& Testor
\cite{hey85}).  They also showed that the chemical composition of the
object is compatible with that of typical LMC \h2\, regions. Israel \&
Koornneef (\cite{ik88}) detected near-IR molecular hydrogen emission
towards the object, partly shocked and partly radiatively excited.
They also confirmed the high extinction of the object from a
comparison of Br\,$\gamma$\, and \hb\, and estimated that N\,159-5
contributes \ab\,25\% to the total flux of the IRAS source LMC\,1518
(Israel \& Koornneef \cite{ikjhk}). More recently, Comer\'on \& Claes
(\cite{cc}) used ISOCAM to obtain an image of N\,159-5 at 15\,$\mu$m.
Similarly, Hunt \& Whiteoak (\cite{hun}) used the Australia Telescope
Compact Array (ATCA) to obtain the highest angular resolution radio
continuum observations in existence of N\,159-5 (beam
8\frac.3\,\x\,7\frac.4).  However, none of these observations were
able to resolve N\,159-5.

\begin{figure*}
\resizebox{\hsize}{!}{\includegraphics{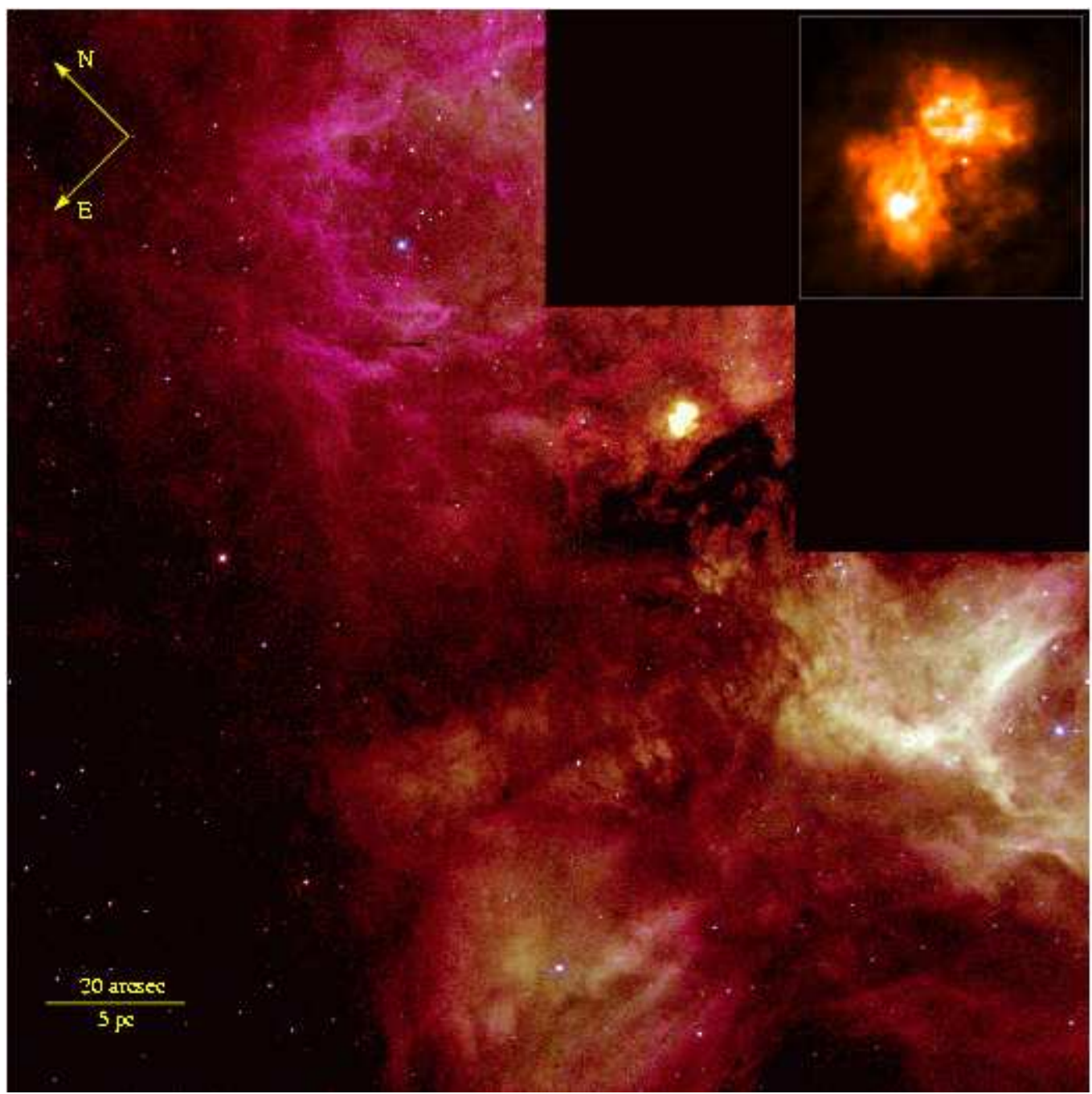}}
\caption{A ``true color'' composite image of the
LMC \h2\, region N\,159  as seen by WFPC2, based on
images taken with filters \ha\, (red), \oiii\, (green), and \hb\, (blue). 
The N\,159-5 ``blob'' is the high excitation compact object lying in the
center of the smaller PC frame.  Note how the hardness of the
radiation field varies from north to south across the image.  The
yellowish colors represent higher excitation gas as traced by the
\oiii\, emission line and are mainly found in the giant ridge in the
southern part of N\,159 and in the blob. In the northern part the
emission is principally due to relatively lower energy \ha\, photons. 
{\bf inset}:A false color image of N\,159-5 in \ha\,
showing the internal morphology of the butterfly-shaped
\h2\, region. Field size \ab\,5\frac\,\x\,5\frac\, (1.3\,\x\,1.3 pc).
}
\label{mosaic}
\end{figure*}

\section{Observations}

The observations of N\,159-5 described in this paper were obtained
with the Wide Field Planetary Camera (WFPC2) on board the {\it HST} on
September 5, 1998 as part of the project GO\,6535.  We used several
wide- and narrow-band filters (F300W, F467M, F410M, F547M, F469N,
F487N, F502N, F656N, F814W) to image the stellar population as well as
the ionized gas. The observational techniques, exposure times, and
reduction procedures are similar to those explained in detail in Paper
I. 

\section{Results}
 
\subsection{Morphology}

In Fig.\,\ref{mosaic} we present the WFPC2 image of the eastern part
of the giant \h2\, region N\,159. This image reveals a very turbulent
medium in which ionized subarcsecond structures are interwoven with
fine absorption features.  A large number of filaments, arcs, ridges,
and fronts are clearly visible. In the south-western part of
Fig.\,\ref{mosaic}, we note a relatively large, high excitation ridge
bordering a remarkable cavity \ab\,25\,\frac\, (\ab\,6\,pc) in size.
Another conspicuous cavity lies in the northern part of the image.
These are most probably created by strong winds of massive stars.
Moreover, a salient, dark gulf running westward into N\,159 cuts the
glowing gas in that direction and as it advances takes a filamentary
appearance. A comparison with the CO map of Johansson et al.
(\cite{joh}) indicates that this absorption is due to the molecular
cloud N\,159-E.

The \h2\, blob N\,159-5 stands out as a prominent high excitation
compact nebula in the center of the WFPC2 field (Fig.\,\ref{mosaic}), at
the edge of two distinct absorption lanes of size
\ab\,3\frac\,\x\,13\frac\, (\ab\,0.8\,\x\,3.3 pc).  

The most important result of our WFPC2 observations is shown in the
inset of Fig.\,\ref{mosaic}, namely the N159-5 blob resolved for the
first time. In fact N\,159-5 consists of two distinct ionized
components separated by a low brightness zone the eastern border of
which has a sharp front.  The overall shape of N\,159-5 is reminiscent
of a butterfly or {\it papillon} in French.\footnote{The term
``butterfly'' is already used to designate several planetary nebulae
in our Galaxy: M\,76, M2-9, NGC\,6302, NGC\,2440, and
PN\,G010.8+18.0. It has also recently been used to describe the K-L
nebula (see Sect. 4).}  The centers of the two wings are
\ab\,2\frac.3 (0.6 pc) apart.  The brightest part of
the right wing appears as a ``smoke ring'' or a doughnut with a
projected radius of \ab\,0\frac.6 (0.14 pc). The left wing is
characterized by a very bright ``globule'' of radius \ab\,0\frac.4
(0.1 pc) to which are linked several bright stripes all parallel and
directed towards the central sharp front.

An obvious questions is: where is (are) the ionizing star(s) of
N159-5? No conspicuous stars can be detected within the Papillon
itself, although its overall high excitation and morphology require
the source of ionization to be very close to the center of this
structure. A faint star of $y$\,=\,17.9 mag can be seen between the
two wings (Fig.\,\ref{mosaic}, inset) and may well be the major source
of ionization, heavily obscured by foreground dust. At least three
more stars weaker than $y$\,\ab\,20 mag are detected (not visible in
Fig.\,\ref{mosaic}), two of them lying in the brightest parts of the
smoke ring and one towards the other wing, east of the front. No star
is detected towards the globule.

\subsection {Nebular reddening}

The {\it HST} observations allow us to study  the spatial variation
of the extinction in the direction of the Papillon nebula.  The
\ha/\hb\, ratio is high in both wings (Fig.\,\ref{rapport}a), varying
between 5 and 10, corresponding to a visual extinction $A_{V}$ between
1.5 and 3.5 mag, if the LMC interstellar reddening law is
used (Pr\'evot et al. \cite{pre}). The extinction towards the zone
separating the two wings also shows comparable ratios.  The
\ha/\hb\, map was used to de-redden the \hb\, flux on a pixel 
to pixel basis.

\subsection{Ionized gas emission}

\begin{figure}
\resizebox{\hsize}{!}{\includegraphics{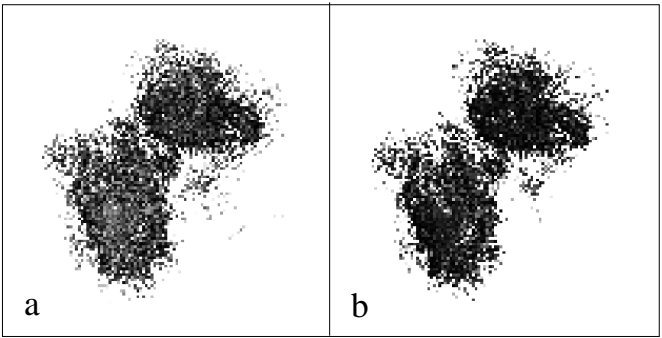}}
\caption{Spatial variation of the extinction and ionization 
across N\,159-5.  a) The \ha/\hb\, ratio  varies between 5 and
10 corresponding to a visual extinction between 1.5 and 3.5 mag.  b)
The \oiii\,\lam\,5007/\hb\, varies between 3.5 and 8.  The field
size and orientation of both frames are identical to the inset of
Fig.\,\ref{mosaic}. }
\label{rapport}
\end{figure}

The \oiii\,\lam\,5007/\hb\, map also displays the  butterfly-like
structure (Fig.\,\ref{rapport}b) with line ratios varying 
between 3.5 and 8.  The band separating the wings has overall smaller
values. A comparison of the \oiii\, and \ha\, images shows that the
Papillon has the same size and morphology in both filters. This
suggests a hard radiation field in which the high excitation O$^{++}$
ions occupy the same zone as the ionized hydrogen. 
A simple calculation (Paper II) shows that almost all
oxygen atoms are doubly ionized.

We measure a total \hb\, flux F(\hb) = 2.68\,\x\,10$^{-13}$ erg
\cm2\,\sec\, above 3$\sigma$ level for both wings (accurate to
\ab\,3\%).  Correcting for the extinction (Sect. 3.2) gives
$F_{0}$\,=\,5.35\,\x\,10$^{-12}$ erg \cm2\,\sec.  The flux is not
equally distributed between the two wings since
\ab\,60\% is generated by the eastern wing (globule), 
whereas the western wing (smoke ring) contributes by \ab\,40\%. 

A Lyman continuum flux of $N_{L}$\,=\,4.17\,\x\,10$^{48}$ photons
\sec\, can be derived for the whole N\,159-5 taking
$T_{e}$\,=\,10\,500\,K (Heydari-Malayeri \& Testor \cite{hey85}), 
assuming a distance of 55 kpc, and considering that the \h2\, region
is ionization-bounded. A single O8V star can account for this ionizing
flux (Vacca et al. \cite{vac}, Schaerer \& de Koter
\cite{sch}). However, this should be viewed as a lower limit, 
since the region is probably not ionization-bounded.

\section{Discussion and concluding remarks}

The three Magellanic Cloud blobs studied so far with {\it HST} (N\,81,
N\,88A, and N\,159-5) represent very young massive stars leaving their
natal molecular clouds. While N\,81 is a rather isolated starburst,
N\,88A and N\,159-5 are formed in richer regions of gas where a
preceding generation of massive stars has taken place.  Being very
young, these two regions are also similar in that they are heavily
affected by dust.  In spite of these similarities, N\,159-5 has a
bewildering morphology which is seen neither in the other blobs, nor,
generally speaking, in any of the known Magellanic Cloud \h2\,
regions. The global morphology of the Papillon and more especially the
presence of peculiar fine structure features in the wings make this
object a unique \h2\, region in the LMC. It is also the first bipolar
nebula indicating young massive star formation in an outer galaxy.  In
this respect the Papillon may even represent a new type of very young
\h2\, region in the Magellanic Clouds overlooked so far because of
insufficient spatial resolution.

Even though high resolution spectroscopy would be necessary for a
conclusive picture of the Papillon's nature, the parallel ray-like
features of the globule as well as the smoke ring strongly suggest a
dynamical origin. Several observational facts also advocate a common
process for the formation of these two compact features: their
proximity (\ab\,2\frac.3), their location in a distinct, more diffuse
nebular structure, and the uniqueness of the phenomenon in the large
field of the N\,159 complex.  In order to explain the observed
morphology of N\,159-5, two different models can be envisaged.

1) We are looking at a bipolar region produced by strong stellar wind
of hot star(s) hidden behind the central absorption zone.  It has been
shown that in hot, rotating stars the mass loss rate is much larger at
the poles than at the equator (Maeder 1999). This model can account
for the high excitation of the two wings, as well as the global
bipolar morphology.  Although we cannot yet firmly advocate a bipolar
phenomenon, we underline the morphological similarity between the
Papillon and the high resolution image of Kleinmann-Low nebula in
Orion recently obtained with the Subaru 8.3\,m telescope at
2.12\,$\mu$m (Subaru team 1999). This image shows a butterfly-shaped
exploding area produced by the wind of a young cluster of stars, among
which IRc2, a particularly active star estimated to have a mass over
30\,\sm.  However, this model does not explain the smoke ring neither
the parallel stripes.

One may also compare N\,159-5 with the well-studied Galactic 
\h2\, region Sh\,106 which has a prominent bipolar shape, 
marvellously shown in an {\it HST} \ha\, image (Bally et al. 
\cite{bal} and references therein). Its exciting star is hidden in the
absorption region between the two lobes where extinction amounts to 20
mag in the visual. The southern lobe is much brighter than the
northern one because it is blueshifted, while the other is expanding
away from the observer.  Also, both lobes get gradually fainter in
their external parts. However, we do not see such global trends in
N\,159-5 and, more importantly, no smoke ring or globule features are
present in Sh\,106.

2) Alternatively, N\,159-5 may represent two close but distinct nebulae
each with its own massive star providing a high excitation for the
ionized gas. In this possibility, the globule may be a bow shock created
by an O star with a powerful stellar wind moving at speeds of \ab\,10
km\,\sec\, through the molecular cloud (models of Van Buren \& Mac Low
\cite{van} and references therein). The presence of the parallel rays
and the sharp front make this explanation attractive.  Since no stars
are detected towards the globule, we may speculate that the star is
hidden behind the bow shock/globule that is heading towards the
observer. As for the smoke ring, it may be due to the interaction of a
stellar wind from a central star with the surrounding medium creating
a bubble structure. If this is the case, the mass loss should be
important, comparable to that observed in Luminous Blue
Variables.  However, LBVs are evolved massive stars and it is not
clear how this phenomenon can occur in such a young region.

Another explanation for the smoke ring can be provided by the
mentioned bow shock models. They predict the formation of a stellar
wind bubble when the star's motion is subsonic. The smoke ring can
therefore be due to such a slower moving star. If this picture is
correct, we are witnessing a very turbulent star forming site where
massive stars formed in group are leaving their birthplace.

One should stress the noteworthy absence of prominent stars towards
the Papillon.  As in the case of the SMC N\,88A (Paper II), this is
certainly due to the very young age and the high dust content of this
star formation region. N\,159-5 is just hatching from its natal
molecular cloud, and its exciting stars should therefore be buried
inside dust/gas concentrations. In order for a star of type O8 with
$M_{V}$\,=\,--4.66 mag (Vacca et al. \cite{vac}) to remain undetected
in our Str\"omgren $y$ image, we need extinctions larger than
$A_{V}$\,=\,6 mag, which is quite possible given the above
estimates. However, N\,159-5 will gradually evolve into a more
extended, less dense region exhibiting its exciting stars, like the
SMC N\,81 (Paper I).

\begin{acknowledgements}
We are grateful to Dr. J. Lequeux for a critical reading of the
manuscript.  We would like also to thank an unknown referee for
suggestions that improved the manuscript.  We are also indebted to
Dr. L.B.E. Johansson for providing us with the CO map of the N\,159
molecular cloud.  VC would like to acknowledge the financial support
from a Marie Curie fellowship (TMR grant ERBFMBICT960967).
\end{acknowledgements}

{}

\end{document}